\documentclass[showpacs,preprintnumbers,amsmath,amssymb]{revtex4}

\usepackage{graphicx}
\usepackage{dcolumn}
\usepackage{bm}

\begin{document}

\preprint{...}

\title{ Topological quantization of self-dual Chern-Simons vortices  on Riemann Surfaces }

\author{Yongqiang Wang$^1$ $^2$}
\thanks{Corresponding author}\email{wyq02@st.lzu.edu.cn}
\author{Yuxiao Liu$^2$}
\author{Zhenhua Zhao$^1$ $^2$}
\author{Yishi Duan$^2$}

\affiliation{
 $^1$Institute of Modern Physics, Chinese Academy of Sciences, Lanzhou
730000, People's Republic of China\\
$^2$Institute of Theoretical Physics, Lanzhou University,
 Lanzhou 730000, People's Republic of China}


\begin{abstract}
The self-duality equations of Chern-Simons Higgs theory in a
background curved spacetime are studied by making use of the
$U(1)$ gauge potential decomposition theory and $\phi$-mapping
method. The special form of the gauge potential decomposition is
obtained directly from the first of the self-duality equations.
Using this decomposition, a rigorous proof of magnetic flux
quantization in background curved spacetime is given. Furthermore,
the precise self-dual vortex equation with topological term is
obtained, in which the topological term has always been ignored.

\end{abstract}
 \maketitle

\section{Introduction}
In recent years, Chern-Simons gauge theories have attracted the
attention in various subjects of both physics and mathematics. One
of interdisciplinary topics attracted attention is so-called
Bogomol'nyi-type vortices and self-dual solutions \cite{HKP, JLW,
Dun}. In Chern-Simons gauge theories coupled to scalar matter
field, a important model is Chern-Simons Higgs model. The progress
to this direction has also been achieved in Chern-Simons Higgs
model coupled to background gravity \cite{Sch,S} and Einstein
gravity \cite{Val}.

    In previous paper about Chern-Simons Higgs theory including gravity,
none can indicate or prove quantization of magnetic flux, and
moreover, the conventional self-dual vortex equation is everywhere
away from the zeros of the scalar field and the equation is not
meaningful at the zeros of the scalar field. In this paper we will
discuss these two questions. Firstly, using Duan's $\phi$-mapping
theory \cite{4,5,6,7,8}, we study the topological inner structure
of Bogomol'nyi self-duality equations and obtain directly one
special form of the gauge potential decomposition. Using this
decomposition, we firstly give a rigorous proof of magnetic flux
quantization in background curved spacetime , and one sees that
the inner structure of this vortex labelled only by the
topological indices of the zero points of the complex scalar
field. Secondly, we obtain the precise scalar field equation with
a topological term, which differs from the conventional equation.

This paper is organized as follows. In the next section, we
briefly review Bogomol'nyi bound of the Chern-Simons Higgs theory
coupled to background gravity and discuss the conventional
self-duality equations, Sec. III presents one special form of the
general $U(1)$ gauge potential decomposition from the first
self-duality equation. In Sec. IV we will give a rigorous proof of
magnetic flux quantization. In Sec. V We then obtain the precise
self-dual vortex equation with topological term. The conclusions
of this paper are given in Sec. VI.

\section{ review of Bogomol'nyi bound of the Chern-Simons Higgs IN BACKGROUND GRAVITY}

In this section we review derivation of so-called Bogomol'nyi
bound of the Chern-Simons Higgs theory coupled to background
gravity. We consider a (2+1)-dimensional space-time $M^{3}$, the
metric $g_{\mu\nu}$ of the (2+1)-dimensional manifold $M^{3}$ is
determined by
\begin{equation}\label{001}
ds^{2}=g_{\mu\nu}dx^\mu
dx^\nu=N^{2}(x^{k})dt^{2}-\gamma_{ij}(x^{k})dx^{i}dx^{j},
\end{equation}
where $\gamma_{ij}$ is the metric of two-dimensional spatial
hypersurface and $i,j,k=1,2$. The Chern-Simons Higgs Lagrangian is
described by
\begin{equation}\label{002}
S=\int d^{3}x\sqrt{g}\left[
\frac{\kappa}{4}\frac{\epsilon^{\mu\nu\lambda}}{\sqrt{g}}
A_{\mu}F_{\nu\lambda} +\frac{1}{2}g^{\mu\nu}D_{\mu}\phi
D_{\nu}\phi^{\ast}-V(\|\phi\|) \right],
\end{equation}
where $\phi$ is the designated charged Higgs complex scalar field
minimally coupled to an Abelian gauge field,
$\frac{\kappa}{4}\frac{\epsilon^{\mu\nu\lambda}}{\sqrt{g}}
A_{\mu}F_{\nu\lambda}$ is the so-called Chern-Simons term in
curved spacetime, and the covariant derivative is
$D_{\mu}\phi=\partial_{\mu}\phi-ieA_{\mu}\phi$. Since the
Bogomol'nyi limit is our interest, the form of the scalar
potential $V(\|\phi\|)$ is taken to be
\begin{equation}\label{003}
V(\|\phi\|)=\frac{e^{4}}{8\kappa^{2}}\|\phi\|^{2}(\|\phi\|^{2}-v^{2})^{2}.
\end{equation}

The energy-momentum tensor is derived as usual
\begin{equation}\label{004}
T_{\mu\nu}=\frac{1}{2}(D_{\mu}\phi^{\ast}D_{\nu}\phi+
D_{\nu}\phi^{\ast}D_{\mu}\phi)-g_{\mu\nu}\left[\frac{1}{2}g^{\rho\sigma}
D_{\rho}\phi^{\ast}D_{\sigma}\phi-V(\|\phi\|)\right].
\end{equation}
An appropriate rearrangement of stress components of the static
energy-momentum tensor gives
\begin{eqnarray}\label{005}
T^{ij}&=&\frac{1}{2}\gamma^{ij}\left[\frac{\kappa^{2}}{2e^{2}}
\frac{B^{2}}{\|\phi\|^{2}}-V(\|\phi\|)\right]-\frac{1}{2}(\gamma^{ij}\gamma^{kl}
-\gamma^{ik}\gamma^{jl}-\gamma^{il}\gamma^{jk})D_{k}\phi^{\ast}
D_{l}\phi \nonumber
 \\
&=&\frac{\kappa^{2}}{2e^{2}}\frac{\gamma^{ij}}{\|\phi\|^{2}}
\left[B-\frac{e^{3}}{2\kappa^{2}}\|\phi\|^{2}(\|\phi\|^{2}-v^{2})\right]
\left[B+\frac{e^{3}}{2\kappa^{2}}\|\phi\|^{2}(\|\phi\|^{2}-v^{2})\right]\nonumber\\
&&+\frac{1}{8}\left\{ \left[\left({D^{i}\phi\mp
i\frac{\epsilon^{ik}}{\sqrt{\gamma}} \gamma_{kl}D^{l}
\phi}\right)^{\ast} \left(D^{j}\phi\pm
i\frac{\epsilon^{jm}}{\sqrt{\gamma}}
\gamma_{mn}D^{n} \phi\right) \right. \right.\nonumber\\
&&\hspace{14mm} \left. +\left({D^{j}\phi\pm
i\frac{\epsilon^{jk}}{\sqrt{\gamma}} \gamma_{kl}D^{l}
\phi}\right)^{\ast} \left(D^{i}\phi\mp
i\frac{\epsilon^{im}}{\sqrt{\gamma}}
\gamma_{mn}D^{n} \phi\right) \right] \nonumber\\
&&\hspace{6mm} +\left[\left({D^{i}\phi\pm
i\frac{\epsilon^{ik}}{\sqrt{\gamma}} \gamma_{kl}D^{l}
\phi}\right)^{\ast} \left(D^{j}\phi\mp
i\frac{\epsilon^{jm}}{\sqrt{\gamma}}
\gamma_{mn}D^{n} \phi\right) \right.\nonumber\\
&&\hspace{14mm}\left.\left. + \left. \right.\left({D^{j}\phi\mp
i\frac{\epsilon^{jk}}{\sqrt{\gamma}} \gamma_{kl}D^{l}
\phi}\right)^{\ast} \left(D^{i}\phi\pm
i\frac{\epsilon^{im}}{\sqrt{\gamma}} \gamma_{mn}D^{n} \phi\right)
\right]\right\}  \label{tij},
\end{eqnarray}
where $\gamma^{ij}$ is inverse of the $\gamma_{ij}$,
$\sqrt{\gamma}=\sqrt{\det \gamma_{ij}}$, and the magnetic field is
defined by
$B=-\frac{\epsilon^{ij}}{\sqrt{\gamma}}\partial_{i}A_{j}$.

We obtain the first-order Bogomol'nyi equations from
Eq.~(\ref{tij})
\begin{eqnarray}\label{006}
D_{i}\phi\pm
i\sqrt{\gamma}\epsilon_{ij}\gamma^{jk}D_{k}\phi=0,\nonumber
\\B=\pm\frac{e^{3}}{2\kappa^{2}}\|\phi\|^{2}(\|\phi\|^{2}-v^{2}).
\end{eqnarray}

When the scalar field $\phi$ is decomposed into its phase and
magnitude: $\phi=\rho^{\frac{1}{2}}e^{iw}$, the first of the
self-duality equations (\ref{006}) determines the gauge field:
\begin{equation}\label{007}
A_{i}=-\frac{1}{e}\partial_{i}w\pm\frac{1}{2e}\sqrt{\gamma}\epsilon_{ij}\gamma^{jk}\partial_{k}\ln\rho.
\end{equation}
We can see that the first equation expresses the spatial
components of the gauge field $A_i$ in terms of the scalar field.
Substituting it into the second Bogomol'nyi equation (\ref{006}),
the second equation reduces to a nonlinear elliptic equation for
the scalar field density $\rho$:
\begin{equation}\label{008}
\Box\ln\rho=\frac{e^{4}}{\kappa^{2}}\rho(\rho-\nu^{2}),
\end{equation}
where
$\Box=\frac{1}{\sqrt{\gamma}}\partial_{i}(\sqrt{\gamma}\gamma^{ij}\partial_{j}\ln(\phi\phi^{\ast}))$
is the Laplacian and this equation is not solvable, or even
integrable.

\section{U(1) gauge potential decomposition of self-duality equations}
It is well known that the complex scalar field $\phi$ can be
looked upon as a section of a complex line bundle with base
manifold $M$ \cite{Dubrovin}. Denoting  the charged Higgs complex
scalar field $\phi$ as
\begin{equation}\label{009}
\phi=\phi^{1}+i\phi^{2},
\end{equation}
where $\phi^{a}(a=1,2)$ are two components of a two-dimensional
vector field $\vec{\phi}=(\phi^{1},\phi^{2})$ over the base space,
one can introduce the two-dimensional unit vector
\begin{equation}\label{010}
n^{a}=\frac{\phi^{a}}{\|\phi\|},\;\;\;\|\phi\|=(\phi\phi^{\ast})^{\frac{1}{2}}.
\end{equation}

Let us consider the first of self-duality equations (\ref{006})
(firstly, we choose the upper signs):
\begin{equation}\label{011}
D_{i}\phi+i\sqrt{\gamma}\epsilon_{ij}\gamma^{jk}D_{k}\phi=0,
\end{equation}
or,
\begin{equation}\label{012}
\left\{\begin{array}{ll}
D_{1}\phi+i\sqrt{\gamma}\gamma^{21}D_{1}\phi+ i\sqrt{\gamma}\gamma^{22}D_{2}\phi=0,\\
D_{2}\phi-i\sqrt{\gamma}\gamma^{11}D_{1}\phi-
i\sqrt{\gamma}\gamma^{12}D_{2}\phi=0.
\end{array}
\right.
\end{equation}
Substituting Eq. (\ref{009}) into the first of the above equations
, we obtain two equations
\begin{eqnarray}\label{013}
\partial_{1}\phi^{1}-\sqrt{\gamma}(\gamma^{21}\partial_{1}\phi^{2}+\gamma^{22}\partial_{2}\phi^{2})+eA_{1}\phi^{2}+\sqrt{\gamma}(\gamma^{21}eA_{1}\phi^{1}+\gamma^{22}eA_{2}\phi^{1})=0, \nonumber\\
\partial_{1}\phi^{2}+\sqrt{\gamma}(\gamma^{21}\partial_{1}\phi^{1}+\gamma^{22}\partial_{2}\phi^{1})-eA_{1}\phi^{1}+\sqrt{\gamma}(\gamma^{21}eA_{1}\phi^{2}+\gamma^{22}eA_{2}\phi^{2})=0.
\end{eqnarray}
Making use of the above relations, we derive:
\begin{equation}\label{014}
\partial_{1}\phi^{\ast}\phi-\partial_{1}\phi\phi^{\ast}=-2ieA_{1}\|\phi\|^{2}+i\sqrt{\gamma}\gamma^{22}(\partial_{2}\phi^{\ast}\phi+\partial_{2}\phi\phi^{\ast})+i\sqrt{\gamma}\gamma^{21}(\partial_{1}\phi^{\ast}\phi+\partial_{1}\phi\phi^{\ast}),
\end{equation}
\begin{equation}\label{015}
\sqrt{\gamma}\gamma^{22}(\partial_{2}\phi^{\ast}\phi-\partial_{2}\phi\phi^{\ast})=-2ie\sqrt{\gamma}\gamma^{22}A_{2}\|\phi\|^{2}-i(\partial_{1}\phi^{\ast}\phi+\partial_{1}\phi\phi^{\ast})-i\gamma\gamma^{21}\gamma^{22}(\partial_{2}\phi^{\ast}\phi+\partial_{2}\phi\phi^{\ast})-i\gamma\gamma^{21}\gamma^{21}(\partial_{1}\phi^{\ast}\phi+\partial_{1}\phi\phi^{\ast}).
\end{equation}
To proceed, we need a fundamental identity-one that appear many
 times throughout our study in the gauge
potential decomposition theory:
\begin{equation}\label{016}
\epsilon_{ab}n^{a}\partial_{i}n^{b}=\frac{1}{2i}\frac{1}{\phi^{\ast}\phi}(\partial_{i}\phi^{\ast}\phi-\partial_{i}\phi\phi^{\ast}),
\end{equation}
using this identity, Eq. (\ref{014}) and Eq. (\ref{015}) become
\begin{equation}\label{017}
eA_{1}=-\epsilon_{ab}n^{a}\partial_{1}n^{b}+\frac{1}{2}\sqrt{\gamma}\gamma^{22}\partial_{2}\ln(\phi\phi^{\ast})+\frac{1}{2}\sqrt{\gamma}\gamma^{21}\partial_{1}\ln(\phi\phi^{\ast}),
\end{equation}
\begin{equation}\label{018}
eA_{2}=-\epsilon_{ab}n^{a}\partial_{2}n^{b}-\frac{1}{2}\frac{1}{\sqrt{\gamma}\gamma^{22}}\partial_{1}\ln(\phi\phi^{\ast})-\frac{1}{2}\sqrt{\gamma}\gamma^{21}\partial_{2}\ln(\phi\phi^{\ast})-\frac{1}{2}\frac{\sqrt{\gamma}\gamma^{21}\gamma^{21}}{\gamma^{22}}\partial_{1}\ln(\phi\phi^{\ast}).
\end{equation}
Making use of the relation: $det(\gamma^{ij})det(\gamma_{jk})=1$,
we can obtain a fundamental identity:
\begin{equation}\label{019}
-\frac{1}{2}\frac{1}{\sqrt{\gamma}\gamma^{22}}\partial_{1}\ln(\phi\phi^{\ast})-\frac{1}{2}\sqrt{\gamma}\gamma^{21}\partial_{2}\ln(\phi\phi^{\ast})-\frac{1}{2}\frac{\sqrt{\gamma}\gamma^{21}\gamma^{21}}{\gamma^{22}}\partial_{1}\ln(\phi\phi^{\ast})=-\frac{1}{2}\sqrt{\gamma}\gamma^{12}\partial_{2}\ln(\phi\phi^{\ast})-\frac{1}{2}\sqrt{\gamma}\gamma^{11}\partial_{1}\ln(\phi\phi^{\ast}).
\end{equation}
Eqs. (\ref{011}) and (\ref{012}) can be rewritten as:
\begin{equation}\label{020}
eA_{i}=-\epsilon_{ab}n^{a}\partial_{i}n^{b}+\frac{1}{2}\epsilon_{ij}\sqrt{\gamma}\gamma^{jk}\partial_{k}\ln(\phi\phi^{\ast}).
\end{equation}
From the second of the  equations (\ref{012}), we also obtain the
same conclusion.

Following the same discussion, we obtain the similar conclusion
from
$D_{i}\phi-i\sqrt{\gamma}\epsilon_{ij}\gamma^{jk}D_{k}\phi=0$:
\begin{equation}\label{021}
eA_{i}=-\epsilon_{ab}n^{a}\partial_{i}n^{b}-\frac{1}{2}\epsilon_{ij}\sqrt{\gamma}\gamma^{jk}\partial_{k}\ln(\phi\phi^{\ast}).
\end{equation}

So, from the first self-duality equation $D_{i}\phi\pm
i\sqrt{\gamma}\epsilon_{ij}\gamma^{jk}D_{k}\phi=0$, we get
\begin{equation}\label{022}
A_{i}=-\frac{1}{e}\epsilon_{ab}n^{a}\partial_{i}n^{b}\pm\frac{1}{2e}
\epsilon_{ij}\sqrt{\gamma}\gamma^{jk}\partial_{k}\ln(\phi\phi^{\ast}).
\end{equation}
It is obvious to see that we obtain one special form of the
general $U(1)$ gauge potential decomposition in background curved
spacetime.

As one has showed in \cite{8}, the $U(1)$ gauge potential in flat
space can be decomposed by the Higgs complex scalar field $\phi$
as
\begin{equation}\label{023}
A_{i}=\beta\epsilon_{ab}\partial_{i}n^{a}n^{b}+\partial_{i}\lambda,
\end{equation}
in which $\beta$ is a constant and $\lambda$ is a phase factor.
Comparing Eq. (\ref{022}) with Eq. (\ref{023}), we find that the
term $\frac{1}{2e}
\epsilon_{ij}\sqrt{\gamma}\gamma^{jk}\partial_{k}\ln(\phi\phi^{\ast})$
on the RHS of Eq. (\ref{022}) can not be expressed with the form
of partial derivative, so, this term is not a phase factor
denoting the $U(1)$ transformation in curved space.

\section{Topological quantization of self-dual Chern-Simons vortices}
Based on the decomposition of the gauge potential $A_{i}$
discussed in section III, in the following, we will immediately
give a rigorous proof of magnetic flux quantization in background
curved spacetime. Using the two-dimensional unit vector field
(\ref{010}), we can construct a topological current in curved
spacetime:
\begin{equation}\label{024}
J^{\mu}=-\frac{1}{\sqrt{\gamma}}\epsilon^{\mu\nu\lambda}\partial_{\nu}A_{\lambda}=\frac{1}{e}\frac{1}{\sqrt{\gamma}}\epsilon^{\mu\nu\lambda}\epsilon_{ab}\partial_{\nu}n^{a}\partial_{\lambda}n^{b},
\end{equation}
which is the special case of the general $\phi$-mapping
topological current theory \cite{4}. Obviously, the current
(\ref{024}) is conserved. Following the $\phi$-mapping theory, it
can be rigorously proved that
\begin{equation}\label{025}
J^{\mu}=\frac{2\pi}{e}\frac{1}{\sqrt{\gamma}}\delta^{2}(\vec{\phi})D^{\mu}(\frac{\phi}{x}),
\end{equation}
where
$D^{\mu}(\frac{\phi}{x})=\frac{1}{2}\epsilon^{\mu\nu\lambda}\epsilon_{ab}\partial_{\nu}\phi^{a}\partial_{\lambda}\phi^{b}$
is the vector Jacobians. This expression provides an important
conclusion: $J^{\mu}=0, \; iff \; \vec{\phi}\neq0; \;
J^{\mu}\neq0, \; iff \; \vec{\phi}=0$. Suppose that the vector
field $\vec{\phi}(\phi^{1},\phi^{2})$ possesses $l$ zeros, denoted
as $z_{i}(i=1,...,l)$. According to the implicit function theorem
\cite{9}, when the zero points $\vec{z_{i}}$ are the regular
points of $\vec{\phi}$, that requires the Jacobians determinant
\begin{equation}\label{026}
D\left(\frac{\phi}{x}\right)\bigg\vert_{z_{i}}\equiv
D^{0}\left(\frac{\phi}{x}\right)\bigg\vert_{z_{i}}\neq{0}.
\end{equation}
The solutions of $\vec{\phi}(\phi^{1},\phi^{2})=0$ can be
generally obtained: $\vec{x}=\vec{z_{i}}(t), \; i=1,2, \cdots , l,
\; x^{0}=t.$  It is easy to prove that
\begin{equation}\label{027}
D^{\mu}\left(\frac{\phi}{x}\right)\bigg\vert_{z_{i}}=D\left(\frac{\phi}{x}\right)\bigg\vert_{z_{i}}\frac{dx^{\mu}}{dt}.
\end{equation}
According to the $\delta$-function theory \cite{10} and  the
$\phi$-mapping theory, one can prove that
\begin{equation}\label{028}
J^{\mu}=\frac{2\pi}{e}\frac{1}{\sqrt{\gamma}}\sum_{i=1}^{l}\beta_{i}\eta_{i}\delta^{2}(\vec{x}-\vec{z}_{i})\frac{dx^{\mu}}{dt}\bigg\vert_{z_{i}},
\end{equation}
in which the positive integer $\beta_{i}$ is the Hopf index and
$\eta_{i}= sgn(D(\phi/{x})_{z_{i}})=\pm1$ is the Brouwer degree
\cite{11,5}. Then the density of topological charge can be
expressed as
\begin{equation}\label{029}
J^{0}=\frac{2\pi}{e}\frac{1}{\sqrt{\gamma}}\sum_{i=1}^{l}\beta_{i}\eta_{i}\delta^{2}(\vec{x}-\vec{z}_{i}).
\end{equation}
From Eq. (\ref{024}), it is easy to see that
\begin{equation}\label{030}
J^{0}=-\frac{\epsilon^{ij}}{\sqrt{\gamma}}\partial_{i}A_{j}=B.
\end{equation}
So, the total charge of the system  can be written as
\begin{equation}\label{031}
\Phi=\int B \frac{1}{\sqrt{\gamma}}dx^{2}=\int
J^{0}\frac{1}{\sqrt{\gamma}}dx^{2}=\Phi_{0}\sum_{i=1}^{l}\beta_{i}\eta_{i},
\end{equation}
where $\Phi_{0}=\frac{2\pi}{e}$ is the unit magnetic flux in
curved spacetime, and the topological index $n$ in Eq. (\ref{031})
has the following expression
\begin{equation}\label{032}
n=\sum_{i=1}^{l}\beta_{i}\eta_{i}.
\end{equation}
It is obvious to see that there exist $l$ isolated vortices in
which the $i$th vortex possesses charge
$\frac{2\pi}{e}\beta_{i}\eta_{i}$. The vortex corresponds to
$\eta_{i}=+1$, while the antivortex corresponds to $\eta_{i}=-1$.
One can conclude that vortex configuration given in Eq.
(\ref{031}) is a multivortex solution which possesses the inner
structure described by expression (\ref{032}).

\section{self-dual equation with topological term}

In Sec. II the second Bogomol'nyi self-duality equation
(\ref{008}) is meaningless, when the field $\phi=0$. Moreover, no
exact solutions are known for this equation. In this section,
based on the decomposition of $U(1)$, we obtain the precise
self-dual equation with topological term.

Firstly, based on the special form of the general $U(1)$
decomposition of gauge potential (\ref{022}), we get:
\begin{equation}\label{033}
B=\frac{1}{e}\frac{1}{\sqrt{\gamma}}\epsilon^{ij}\epsilon_{ab}\partial_{i}n^{a}\partial_{j}n^{b}\mp\frac{1}{2e}\epsilon^{ij}\epsilon_{jk}\frac{1}{\sqrt{\gamma}}\partial_{i}(\sqrt{\gamma}\gamma^{kl}\partial_{l}\ln(\phi\phi^{\ast})).
\end{equation}
Substituting above equation into the second Bogomol'nyi
self-duality equation (\ref{006}), we obtain:
\begin{equation}\label{034}
\frac{1}{2}\Box
\ln(\phi\phi^{\ast})=\frac{e^{4}}{2\kappa^{2}}\|\phi\|^{2}(\|\phi\|^{2}-\nu^{2})\mp\frac{1}{\sqrt{\gamma}}\epsilon^{ij}\epsilon_{ab}\partial_{i}n^{a}\partial_{j}n^{b}.
\end{equation}
From Eqs. (\ref{024}) and (\ref{025}), we obtain the
$\delta$-function form of topological term
\begin{equation}\label{035}
\frac{1}{\sqrt{\gamma}}\epsilon^{ij}\epsilon_{ab}\partial_{i}n^{a}\partial_{j}n^{b}=eJ^{0}=\frac{2\pi}{\sqrt{\gamma}}\delta^{2}(\vec{\phi})D(\frac{\phi}{x}).
\end{equation}
The second self-dual equation in Eq. (\ref{006}) then can reduce
to a nonlinear elliptic equation for the scalar field density
($\rho=\phi\phi^{\ast}$)
\begin{equation}\label{036}
\Box\ln\rho=\frac{e^{4}}{\kappa^{2}}\rho(\rho-\nu^{2})\mp4\pi\frac{1}{\sqrt{\gamma}}\delta^{2}(\vec{\phi})D(\frac{\phi}{x}).
\end{equation}
where
$\mp4\pi\frac{1}{\sqrt{\gamma}}\delta^{2}(\vec{\phi})D(\frac{\phi}{x})$
is the density of topological charge of the vortices.

Comparing Eq. (\ref{036}) with Eq. (\ref{008}), one can find that
the conventional self-dual equation (\ref{008}), in which the
topological term has been ignored, is meaningless when the field
$\phi=0$; we get the self-dual equation with topological term,
which is meaning when the field $\phi=0$. Obviously, topological
term is very important to the inner topological structure of the
self-duality equations.
\section{conclusion}
Our investigation is based on the connection between the self-dual
equation of Chern-Simons Higgs model coupled to background gravity
and the $U(1)$ gauge potential decomposition theory and
$\phi$-mapping theory. First, we directly obtain one special form
of $U(1)$ gauge potential decomposition from self-duality
equations. Making use of the decomposition, we give a rigorous
proof of magnetic flux quantization in background curved
spacetime. Moreover, we obtain the inner topological structure of
the Chern-Simons vortex, the multicharged vortex has been found at
the Jacbian determinate $D(\phi/x)\neq0$. It is also showed that
the charge of the vortices is determined by Hopf indices and
Brouwer degrees. Second, we establish the rigorous self-duality
equations with topological term for the first time, in which the
topological term is the density of topological charge of vortex:
$\frac{4\pi}{\sqrt{\gamma}}\delta^{2}(\vec{\phi})D(\frac{\phi}{x})=2eJ^{0}$.
In contrast with the conventional self-duality equation
(\ref{008}), one can see that the self-duality equation with
topological term is valid when the field $\phi=0$; topological
term vanishes and the self-duality equation becomes Eq.
(\ref{008}) when the field $\phi\neq0$.

\section{Acknowledgements}
This work was supported by the National Natural Science Foundation
and the Doctor Education Fund of Educational Department of the
People's Republic of China.

\end{document}